\documentclass[aps,pre,floats,
  twocolumn,showpacs,superscriptaddress]{revtex4-1}

\usepackage{graphicx,epsfig}
\usepackage[english]{babel}
\usepackage[utf8]{inputenc}
\usepackage{graphics,dcolumn,bm,epic, eepic,fleqn,float}
\usepackage{amssymb,amsmath,amsfonts,multirow,rotate,color}
\usepackage{soul}
\usepackage{wasysym}
\usepackage{mathtools}
\newcommand{\sgn}[1]{\operatorname{sgn}\! \left(#1\right)}

\definecolor{Myorange}{cmyk}{0,0.42,1,0}

\newcommand{\avg}[1]{\langle #1 \rangle}

\newcommand{\sign}{\operatorname{sign}}
\newcommand{\lay}[1]{^{[#1]}}

\begin{document}

\title{Interplay between consensus and coherence in a model of
  interacting opinions}

\author{Federico Battiston$^*$}
\affiliation{School of Mathematical Sciences, Queen Mary University
  of London, Mile End Road, E1 4NS, London (UK)}
\author{Andrea Cairoli$^*$}
\affiliation{School of Mathematical Sciences, Queen Mary University
  of London, Mile End Road, E1 4NS, London (UK)}
\author{Vincenzo Nicosia}
\affiliation{School of Mathematical Sciences, Queen Mary University
  of London, Mile End Road, E1 4NS, London (UK)}
\author{Adrian Baule}
\affiliation{School of Mathematical Sciences, Queen Mary University
  of London, Mile End Road, E1 4NS, London (UK)}
\author{Vito Latora}
\affiliation{School of Mathematical Sciences, Queen Mary University
  of London, Mile End Road, E1 4NS, London (UK)}

\begin{abstract}
The formation of agents' opinions in a social system is the
  result of an intricate equilibrium among several driving forces. On
  the one hand, the social pressure exerted by peers favours the
  emergence of local consensus. On the other hand, the concurrent
  participation of agents to discussions on different topics induces
  each agent to develop a coherent set of opinions across all the
  topics in which he is active. Moreover, the prevasive action of
  external stimuli, such as mass media, pulls the entire population
  towards a specific configuration of opinions on different topics.
Here we propose a model in which agents with interrelated
  opinions, interacting on several layers representing different
topics, tend to spread their own ideas to their neighbourhood,
strive to maintain internal coherence, due to the fact that
  each agent identifies meaningful relationships among its opinions on
  the different topics, and are at the same time subject to external
fields, resembling the pressure of mass media. We show that
the presence of heterogeneity in the internal coupling
  assigned by agents to their different opinions allows to obtain
states with mixed levels of consensus, still ensuring that all
  the agents attain a coherent set of opinions.  Furthermore,
  we show that all the observed features of the model are preserved in
  the presence of thermal noise up to a critical temperature, after
  which global consensus is no longer attainable. This suggests the
  relevance of our results for real social systems, where noise is
  inevitably present in the form of information uncertainty and
  misunderstandings. The model also demonstrates how mass media can
be effectively used to favour the propagation of a chosen set of
opinions, thus polarising the consensus of an entire population.
\end{abstract}

\maketitle

\section{Introduction}

The increasing availability of data sets about social relationships,
such as friendship, collaboration, competition, and opinion formation,
has recently spurred a renewed interest for the basic mechanisms
underpinning human dynamics~\cite{lazer09}. Aside with the classical
studies in social sciences and social network
analysis~\cite{Wasserman1994,Scott2000,Jackson2010}, some interesting
contributions to the understanding of social dynamics have lately come
from statistical physics~\cite{castellano2009statistical}, which has
brought in the field new tools and analytical methods to study systems
consisting of many interacting agents. In such wider context, much
effort has been devoted to the study of the dynamics responsible for
opinion formation in populations of interacting agents, and in
particular to a more in-depth understanding of the elementary
mechanism allowing the emergence of global consensus and of the role
of endogenous and exogenous driving forces, including social pressure
and mass media. As a result of this investigation, a plethora of
models of opinion formation have been proposed and
studied \cite{wu1982potts,clifford1973model,holley1975ergodic,
Galam1986,Vazquez2008,Vazquez2008a,axelrod97,galam2002minority}.

Although the majority of those models originally made the simplifying
assumption of considering homogeneous interaction patterns (basically,
regular lattices), the rise of network
science~\cite{strogatz2001exploring,albert2002statistical,newman2003structure,boccaletti2006complex}
provided the tools to overcome this limitation, featuring more
realistic interaction patterns.  More recently, also the role of mass
media in the formation of global consensus has attracted a lot of
interest~\cite{bryant2008media,martins2010mass,carletti06,quattrociocchi2014opinion,gonzalezavella10,gonzalezavella05}.

An aspect of social relationships that has been mostly discarded in
the study of the emergence of consensus is the fact that agents
usually interact in a variety of different contexts, making the
interaction pattern effectively multilayered and multi-faceted.
As a matter of fact, the urge to maintain a certain level of coherence
among opinions on different but related subjects might actually play a
crucial role in determining the reaction of each agent to external
pressure and in facilitating (or hindering) the emergence of global
consensus. Moreover, the balance between the internal tendency towards
coherence and the necessity to adequately respond to social pressure
is naturally dependent on each person's attitude, thus implying a
certain level of heterogeneity. Some individuals may be more prone to
align more closely to the opinions of their neighbors in each of the
different contexts where they interact, putting little or no
importance to the overall coherence of their profile. On the contrary,
some other agents may indeed be more reluctant to change their opinion
on a topic, in spite of being urged by other individuals or media, if
such a change results in a contradiction with another of their
opinions on a different but related subject. 

In this paper we propose a model of opinion formation that takes into
account {\em i)} the concurrent participation of agents to distinct
yet connected interaction levels (representing discussion topics or
social spheres), {\em ii)} the presence of social pressure and {\em
  iii)} the exogenous action of mass media.  Our analysis can be
naturally cast in the framework of multiplex networks
\cite{nicosia2013growing,de2013mathematical,battiston2014structural,Boccaletti2014},
which has recently proven successful for a more realistic modeling of
different social
dynamics~\cite{GomezGardenes2012,Diakonova2014,Quattrociocchi2014,Galam2015,Chmiel2015,Diakonova2015}.
According to this framework, agents are represented by nodes connected
by links of different nature, where links of the same kind belong to
the same layer of the system. Each layer thus represents the
interaction pattern of individuals discussing a given topic. Different
layers are in general endowed with different topologies, to mimic
multi-layer real-world social systems where distinct interaction
patterns are present at different levels. Peer social pressure occurs
on each topic through intra-layer links. The opinions of an individual
on the different topics are also driven towards a specific state by
the tension towards internal agent's coherence, represented by a
preferred configuration of opinions on different topics. Mass media
are introduced as fields acting uniformly on all the agents at the
level of each single topic. 

The resulting model is a natural extension of the traditional Ising model of magnetic interaction~\cite{onsager1944crystal} and of more recent variations introduced to take into account the effect of
external forces on the emergence of consensus~\cite{doyle2014social}, in the spirit of less and more recent work connecting statistical mechanics of disordered systems and opinion dynamics~\cite{galam1,galam2,galam3}. The key ingredient of heterogeneous distributed couplings between opinions lead to interesting equilibrium states, where agents can remain fully coherent while a variable level of global consensus is attained, depending on the strength of the pressure exerted by mass media. This clearly resembles the dynamics observed in real societies, thereby supporting the relevance of our approach.

\begin{figure*}[!htb]
\centering
\includegraphics[scale=0.3]{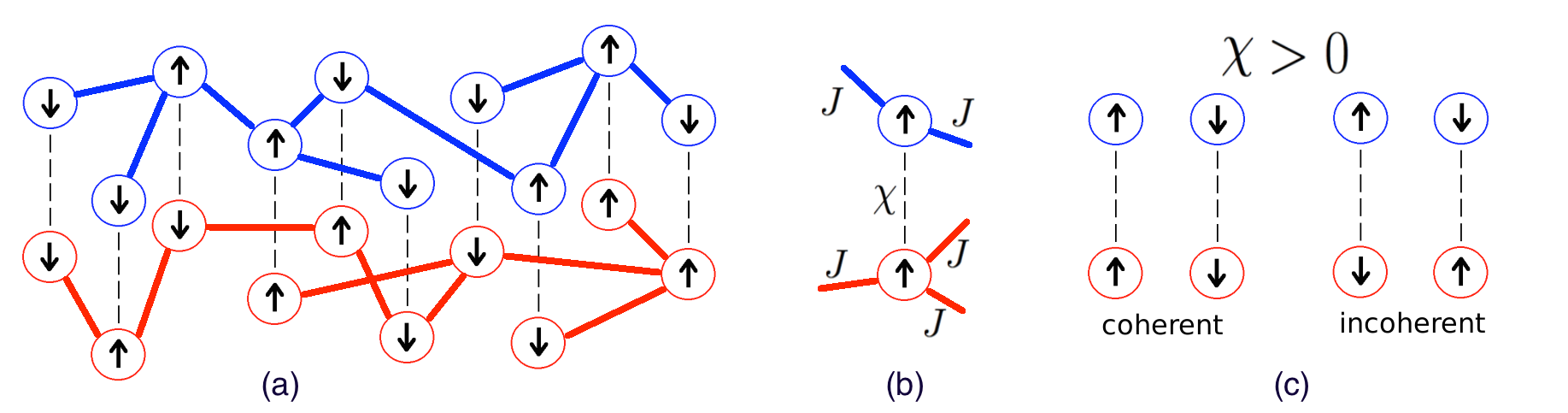}
\caption{(a) Model of interacting opinions for $M=2$ different
 topics. Every agent in the population expresses a binary opinion on
 each subject $\alpha=1,2$ (upward or downward arrows) and may interact
 with other individuals. The pattern of intra-layer interactions
 (blue/red lines for each $\alpha$ respectively) may in general be
 different. (b) Opinions may change according to both social pressure
(weighted by the parameter $J$) and external
 fields, e.g. mass media (not shown). Individuals are also
 differently prone to change one or more of their opinions, according
 to their internal coherence. Such effect is
 taken into account by identifying natural couplings between the
 opinions of an agent at the two layers, weighted by the agent-dependent 
 parameter $\chi_i$. (c) The internal coherence/incoherence of an agent is
 determined by the sign of its opinions (orientation of the arrows):
 when $\chi_i > 0$, coherent configurations are those with both
 opinions of the same sign (left couple), whereas incoherent ones
 present opinions of different sign (right couple). The opposite
 holds if $\chi_i < 0$.}
\label{fig:1}
\end{figure*}

\section{Model}
We consider a population of $N$ individuals interacting
through $M$ different layers, representing different topics or
subjects. The network of each layer $\alpha=1,\ldots,M$ represents the
pattern of interactions among agents on a specific topic, which is in
general distinct from those of the other layers, and is encoded by the
adjacency matrix $A\lay \alpha = \{a_{ij}\lay
\alpha\}_{i,j=1,\ldots,N}$, whose element $a_{ij}\lay{\alpha}=1$ only
if agent $i$ and agent $j$ are neighbors on layer $\alpha$, and equal
to zero otherwise. The structure of the overall interaction pattern is
thus concisely represented by the vector of adjacency matrices
$\mathcal{A} = \{A\lay{1}, \ldots, A\lay{M}\}$, where all the matrices
$A\lay{\alpha}$ are in general distinct.
Each agent $i=1,\ldots,N$ expresses a binary opinion $s_i\lay \alpha =
\pm 1$ on each subject $\alpha=1,\ldots,M$. An example with $M=2$ is
shown in Fig.~\ref{fig:1}(a), where upwards and downwards spins
represent the two possible values of $s_{i}\lay{\alpha}$. We assume
that agent opinions evolve over time due to two concurrent
mechanisms. On the one hand, agents are subject to social pressure
from their peers on each layer (denoted by the red and blue links in Fig.~\ref{fig:1}), so that the opinion of agent $i$ on node $\alpha$ will tend to remain aligned with the opinions of its neighbors on the same
layer. This mechanism, based on the elimination of conflicting
opinions on a microscopic scale, has been widely observed in many
real-world social systems \cite{festinger63}, and is responsible for the attainment of
local consensus on each layer. On the other hand, we assume that the
opinions of agent $i$ at the different layers are not independent from
each other but are instead interacting, so that for each agent there
exists a preferred configuration of opinions at the different layers
which is considered \textit{coherent}. For instance, the political
orientation of a person is often related to his/her ideas about
economy and welfare, so that the emergence of consensus with its
neighbors on one subject should remain coherent with its current
opinions on the other layers.  Moreover, we imagine that agents are
exposed, on each layer, to the action of mass-media, a mean-field
external force which preferentially drives their opinions towards
either $+1$ or $-1$.

We formalize the interplay of these concurrent dynamics by defining
the functional:
\begin{equation}
f_i\lay \alpha = J \sum_{j=1}^N  a_{ij}\lay \alpha s_j\lay \alpha
+h\lay \alpha+\gamma\frac{\chi_i}{J} \sum_{\mathclap{\substack{\beta = 1 \\ \beta \neq \alpha}}}^M s_i\lay \beta
\label{eq:1}
\end{equation}
for each agent $i$ and each topic $\alpha$. The first sum on the rhs
of Eq.~(\ref{eq:1}) represents the social pressure exerted on $i$ by
its neighbors on layer $\alpha$, and is weighted by the
coefficient $J$, which models its intrinsic
permeability to social pressure. 
The variables $h\lay{\alpha}$ represent the external
effect of mass-media on the formation of agents' opinions, which are
considered in this case as a mean-field force acting homogeneously on
all the agents of a layer. Finally, the second sum represents the
tendency of agent $i$ towards internal coherence, where the global
parameter $\gamma$ sets the relative importance of internal coherence
and social pressure. Specifically, when $\gamma \simeq  0$ the opinions of the agents are mainly driven by peer and external pressure, whereas when $\gamma\to\infty$ they are determined by the internal coherence, such that coherent configurations are strongly favoured.

This setup is depicted in Fig.~\ref{fig:1}(b) for the case $M=2$, where links with different colors indicate the connections of an agent at the two layers. In practice, $J$ is the strength of the interaction of each agent with its neighbors, while $\chi_i$ determines the importance (and sign) of internal agent coherence. In this case, as shown in
Fig.~\ref{fig:1}(c), the preferred configuration of agent's $i$ spins
is concordant if $\chi_i>0$ and discordant if $\chi_i<0$. We notice
that the actual value of $\chi_i$, which in the following always
lie in the interval $[-1,1]$, is a measure of how much agent $i$ is
flexible towards a change of one of its opinions, eventually leading
to configurations which do not agree with what it would consider a
coherent configuration of its spins. In other words, agents for which
$|\chi_i|\simeq 0$ assign less importance to internal coherence
and more relevance to social pressure, while the opposite happens when
$|\chi_i| \simeq 1$.

In our model, the opinions of each agent evolve towards configurations
which maximize the function $F_i \lay \alpha = s_i\lay \alpha f_i\lay
\alpha$, in order to attain, at the same time, internal coherence and
local consensus with their neighbors on each layer. As a consequence,
agents will naturally prefer configurations of spins at all layers
which ensure a balanced trade-off between social pressure and
coherence, depending on the respective values of the parameters $\gamma$, $J$, $\chi_i$, and of the external fields $h \lay \alpha$.  Although being a somehow simplified model of real-life
interactions, where not just binary but also intermediate opinions
between two extremes are possible and agents might respond differently
to social pressure and to the external effect of mass-media, this model turns out to be
already general enough to investigate the elementary mechanisms
driving interacting opinions.

Numerical implementation of this dynamical evolution is
obtained through extensive Monte Carlo simulations, adopting an
appropriately modified version of the Glauber
algorithm~\cite{Glauber1963}.  In particular, at each step we update
all the spins $s_i \lay{\alpha}, \> i=1,\ldots, N, \>\alpha=1,2,
\ldots, M$ in a random order. The update is performed by proposing a
flip of the current spin $s\lay{\alpha}_i \rightarrow
-s\lay{\alpha}_i$ and accepting the flip only when the new
configuration leads to a larger value of the function
$F_i\lay\alpha$. Every time a flip is accepted, $F_i\lay\alpha$ is
also updated according to the new configuration. Clearly, the form of
$f_i\lay \alpha$ captures both the contributions of intra-topic and
inter-topic couplings and those of the existing external fields, so
that larger values of $F\lay{\alpha}_i$ correspond to
preferred configurations for node $i$.

Clearly, these rules imply a deterministic evolution of the
  opinions, which is not observed in real social systems. We then need
  to account for the presence of stochastic noise. Its simulation is
  realized by introducing a parameter $T\ge 0$, which may be regarded
  as a social temperature in analogy with magnetic systems, induced by
  all those mechanisms which drive the system out of its deterministic
  dynamics, such as partial information or misunderstandings. We
  include such thermal noise in the dynamics of our model in a
  standard way: when $T > 0$, an agent $i$ may change its opinion on
  the topic $\alpha$ even if it leads to configurations with a smaller
  $F_i \lay\alpha$ with probability $e^{\frac{\Delta F_i
      \lay\alpha}{T}}$, with $\Delta F_i \lay\alpha$ being the
  variation in $F_i \lay\alpha$ due to the flip of thew spin
  $s_{i}\lay{\alpha}$.

We are interested in understanding how the presence of three
concurrent factors, namely the response to social pressure, the
tension towards internal coherence and the presence of an external
mean-field force on each layer, affects the emergence of consensus in
the population. We consider three scenarios, namely {\em i)} the case
in which $\chi_i=1, \forall i$ (homogeneous agents); {\em ii)} the
case in which $\chi_i$ is a random variable sampled from a certain
probability distribution (heterogeneous agents); and finally {\em
  iii)} the case in which the dynamics is affected by noise. For the
sake of simplicity, in the following we focus on the case of two
interconnected layers, i.e. $M=2$, and we set $J=1$, so that
the relative importance of internal coherence and social pressure is
determined, for each agent, by the product $\gamma\chi_i$. The
  model is numerically investigated in the three different setups
  described above, where the parameters $\gamma$, $h\lay{1}$, and
  $h\lay{2}$ play the role of control parameters.

We study the emergence of consensus at each layer $\alpha=1,2$ through
the order parameter:
\begin{equation}
M^{\lay \alpha}=\frac{1}{N}\sum_{i=1}^{N} s_i\lay \alpha,
\label{eq:2}
\end{equation}
which satisfies $-1\le M^{\lay \alpha} \le 1$, with $\left| M^{\lay
 \alpha} \right|$ denoting the strength of consensus and
$\sgn{M^{\lay \alpha}}$ indicating which of the opinion is prevalent
among the population. We also define the average internal coherence of
the agents as follows:
\begin{equation}
C =\frac{1}{N}\sum_{i}^{N} \sgn{\chi_{i}} s_i\lay{1} s_i\lay{2}.
\label{eq:3}
\end{equation}
Notice that, when $\gamma > 0$, $C=+1$ if the two spins of
  each agent are coherent with their preferred configuration, while
  $C=-1$ if they are incoherent for every agent. The opposite holds
  when $\gamma<0$.

An interesting remark is that the global function
$H=-\sum_{\alpha = 1}^{M} \sum_{i=1}^{N} F_i \lay \alpha$ (which we do
not consider in this study) effectively is the Hamiltonian of a
multi-layer Ising model, where the population evolves equivalently
towards configurations that minimise $H$. In this sense, our model can
be considered as a generalization of the coupled Ising model on
lattices~\cite{simon2000coupled}. Following this analogy, we
  note that the order parameters $M \lay \alpha$ can be interpreted as
  the magnetization of the different layers of the system.

\section{Results}
We discuss in this section the transition towards coherence and
consensus and the equilibrium properties of the model, focusing on the
dependence of the order parameters $C$ and $M \lay \alpha$ in
Eqs.~(\ref{eq:2}-\ref{eq:3}) on the parameter $\gamma$ and on the
external fields $h \lay \alpha$. In details, we investigate in Sec.~\ref{SubSec:A} the case of $\chi_i=1$ $\forall i$ and $T=0$, i.e. a population of homogeneous agents in the absence of social noise. In Sec.~\ref{SubSec:B} we consider a population of heterogeneous agents ($\chi_i$  not fixed), while keeping $T=0$. Finally, in Sec.~\ref{SubSec:C} we study the effect of social noise by investigating the dependence on $T$. 

Simulations of the Glauber dynamics described in the previous section are realized by varying the global parameter $\gamma$ adiabatically. The initial configuration is obtained by setting $s_i \lay 1=1$ and $s_i \lay 2=-1$ $\forall i$. We let the system perform two complete hysteresis cycles before recording the resulting configurations. This procedure eliminates possible effects due to the specific initial conditions.

The results presented here are obtained by simulating
the dynamics on a multiplex of two uncorrelated Barabasi-Albert
networks~\cite{sciencebarabasi} with the same average degree
$\left\langle k \right\rangle$=6. Nevertheless, we remark that analogous
qualitative results have been found for different interaction
patterns, such as random graphs with the same density or systems with different values of inter-layer degree-correlation \cite{correlation}, suggesting that the only topological parameter playing a
major role in the long-term behavior of the dynamics is the average
degree of the networks at the two layers.

\begin{figure*}[!tb]
\centering
\includegraphics[width=6.1in]{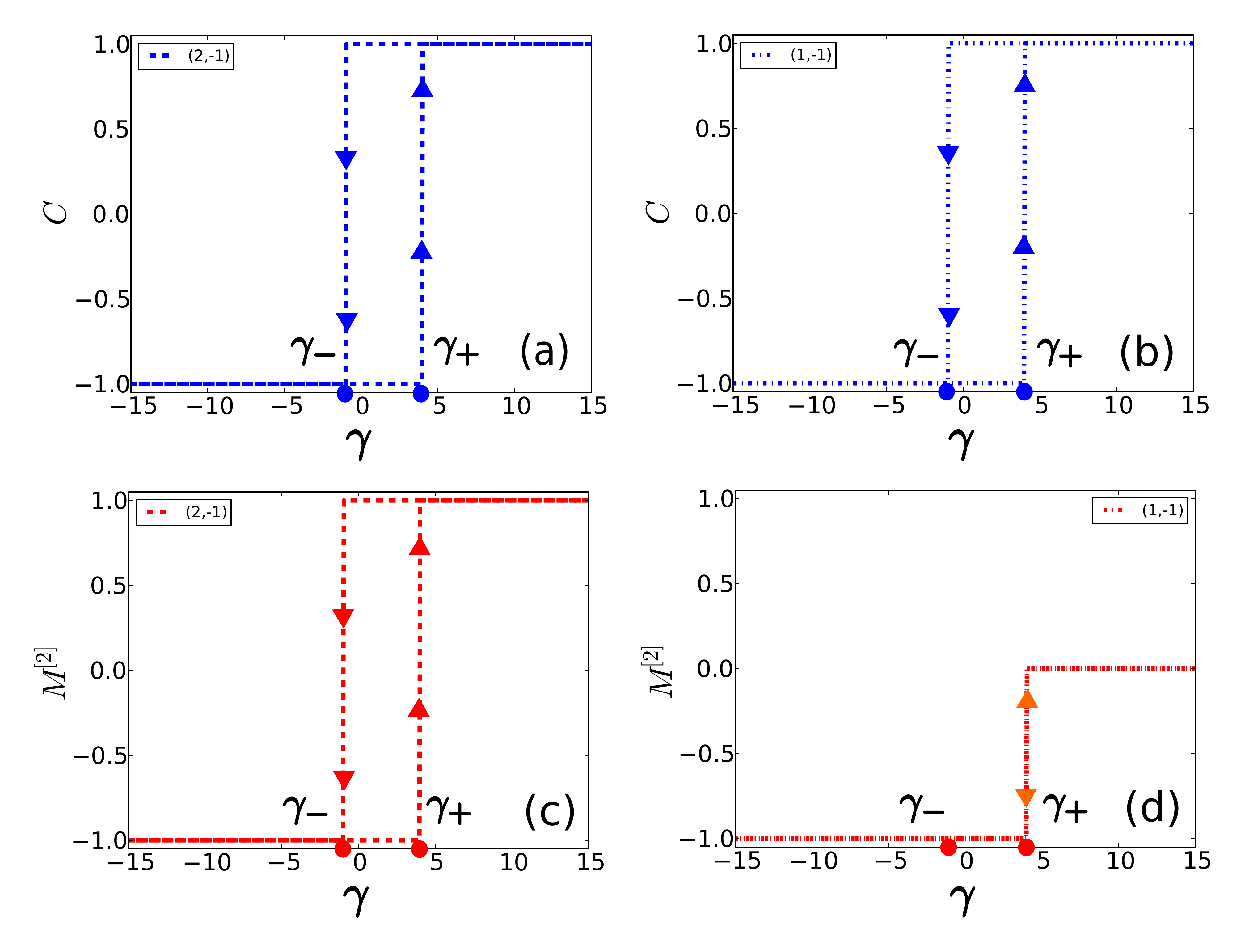}
\caption[]{Values of the average agent coherence $C$ (top
    panels) and of the consensus on the second topic $M\lay 2$ (bottom
    panels) as a function of $\gamma$ for a population of homogeneous
    agents ($\chi_i=1$, $\forall i$). External fields are chosen with
    opposite signs and either different ($|h\lay 1|>|h\lay 2|$) (left
    panels) or equal ($|h\lay{1}| = |h\lay{2}|$) relative intensity
    (right panels). A sharp transition towards full coherence
  ($C=1$), characterised by an hysteresis cycle delimited by
    the transition points $\gamma_+$ and $\gamma_-$, is observed in
  both cases (panels a-b), independently from the relative strength of
  the media. Conversely, the corresponding value of $M\lay 2$ after
  the transition, i.e., when agents are coherent, differs
  significantly: if $|h\lay 1|=|h\lay 2|$ (panel d), we have $M\lay 2
  = 0$, while if $|h\lay 1|>|h\lay 2|$ (panel c), we have $M\lay 2 =
  1$. The presence of a stronger media pressure on a specific topic
  indeed influences also the other one. We note that when
    agents are homogeneous no states of partial consensus are allowed
    on either layer.}
\label{fig:2a}
\end{figure*}

\subsection{\label{SubSec:A}Transition towards full coherence in the case of homogeneous agents}
We consider here the case of homogeneous agents $\chi_i=1$
$\forall i$, in the absence of social noise, i.e., the case $T=0$. The effects induced by the external forces,
e.g., the mass media, are studied by choosing fields with opposite
signs and relative strength according to the two typical cases: $|h\lay1| = |h\lay 2|$ or $|h\lay1|>|h\lay 2|$. We remark that the qualitative behaviour observed does not depend on the specific values of $|h\lay1| $ and $|h\lay 2|$. 
First, we study the transition in coherence as a function of $\gamma$: for fields of both equal and different intensity, we provide evidence of the existence of a sharp transition along with a hysteresis loop. We are also able to propose an empirical relation to estimate the transition points $\gamma_{\pm}$, given the intensity of the fields and the density of the layers. We note that the case of fields with equal signs is somehow trivial, since the opinions on both layers are pulled in the same direction and global consensus emerges easily. Second, we find that a coherent population, i.e. in the regime $\gamma \to \infty$, exhibits either states of full or null consensus and that states of partial consensus cannot be attained in a population of homogenous agents. 

We show examples of the steep transitions that the system
  exhibits by plotting C as a function of $\gamma$ in the top panels
  of Fig.~\ref{fig:2a}(a1) for $|h\lay1|>|h\lay 2|$ and of
  Fig.~\ref{fig:2a}(a2) for $|h\lay1| = |h\lay 2|$. The behavior of
  the coherence is robust with respect to the relative strength of the
  external fields: we always observe a sharp transition from $C=-1$ to
  $C=+1$ characterised by a marked hyseresis loop. However, the actual
  values of $h\lay{1}$ and $h\lay{2}$ deeply affect the corresponding
  level of consensus emerging in the population. This is
  shown in the bottom panels of Figs.~\ref{fig:2a}(a-b), where we plot
  the corresponding value of $M^{\lay 2}$ as a function of $\gamma$.
If the external fields have the same intensity $|h\lay1| = |h\lay 2|$,
we have $M^{\lay 2}=0$ for $\gamma>\gamma_{+}$, while $M^{\lay 2}=-1$
when $\gamma_{-} \leq \gamma \leq \gamma_+$. As the transition
  is sharp, we can always infer the value of $M^{\lay 1}$ from the
  corresponding values of $C$ and $M^{\lay 2}$. In fact, we
  respectively have $M \lay 1=\pm M\lay 2$ when $C=\pm 1$. 

This result has a clear interpretation. When $\gamma$ is increased,
the second term in the rhs of Eq.~(\ref{eq:1}) becomes dominant,
meaning that the agents give more importance to internal coherence
than to social pressure. At the same time, however, none of the two
external fields, which have opposite signs, is able to force a flip of
opinions on the the other layer. This leads naturally to states of
vanishing magnetization on each layer, i.e., no consensus. The
opposite situation is observed when we decrease $\gamma$. Indeed, the
agents become more flexible, so that different opinions on different
topics can coexist. Of course, this tendency gradually increases the
effect of the external fields on their own topic. At the transition
point, the population becomes globally incoherent, whereas the
external fields induce full consensus separately on each layer, with
their sign determining the dominant opinion.

In the case where one of the two fields is larger than the other, i.e.
$|h\lay1| > |h\lay 2|$, the situation is radically different. We
indeed find $M^{\lay 2}=+1$ for $\gamma > \gamma_{+}$, $M^{\lay 2}=-1
$ when $\gamma$ increases in the interval $[\gamma_{-},\gamma_{+}]$,
and $M^{\lay 2}=+1 $ when $\gamma$ decreases in
$[\gamma_{-},\gamma_{+}]$. The interpretation follows
straightforwardly with a reasoning similar to the one reported above
for the case $|h\lay1| = |h\lay 2|$. As $\gamma$ increases, the agents
become more and more inflexible, thus favoring opinions of the same
sign throughout the different topics. Moreover, since $|h \lay 1|$ is
larger than $|h \lay 2|$, states of non-vanishing consensus are
favored. In particular, one of the opinions ends up
prevailing not just on layer $1$ but, through the internal
agent coherence, also on the other layer. Thus, the concurrent effect
of these two mechanisms causes a steep transition towards a state of
both full coherence and full consensus on a single opinion on both the
topics, which is determined by the leading external field. The same
dynamical explanation of the previous case can instead be given for
decreasing values of $\gamma$ beyond $\gamma_{-}$.
\begin{figure*}[!tb]
\centering
\includegraphics[width=6.1in]{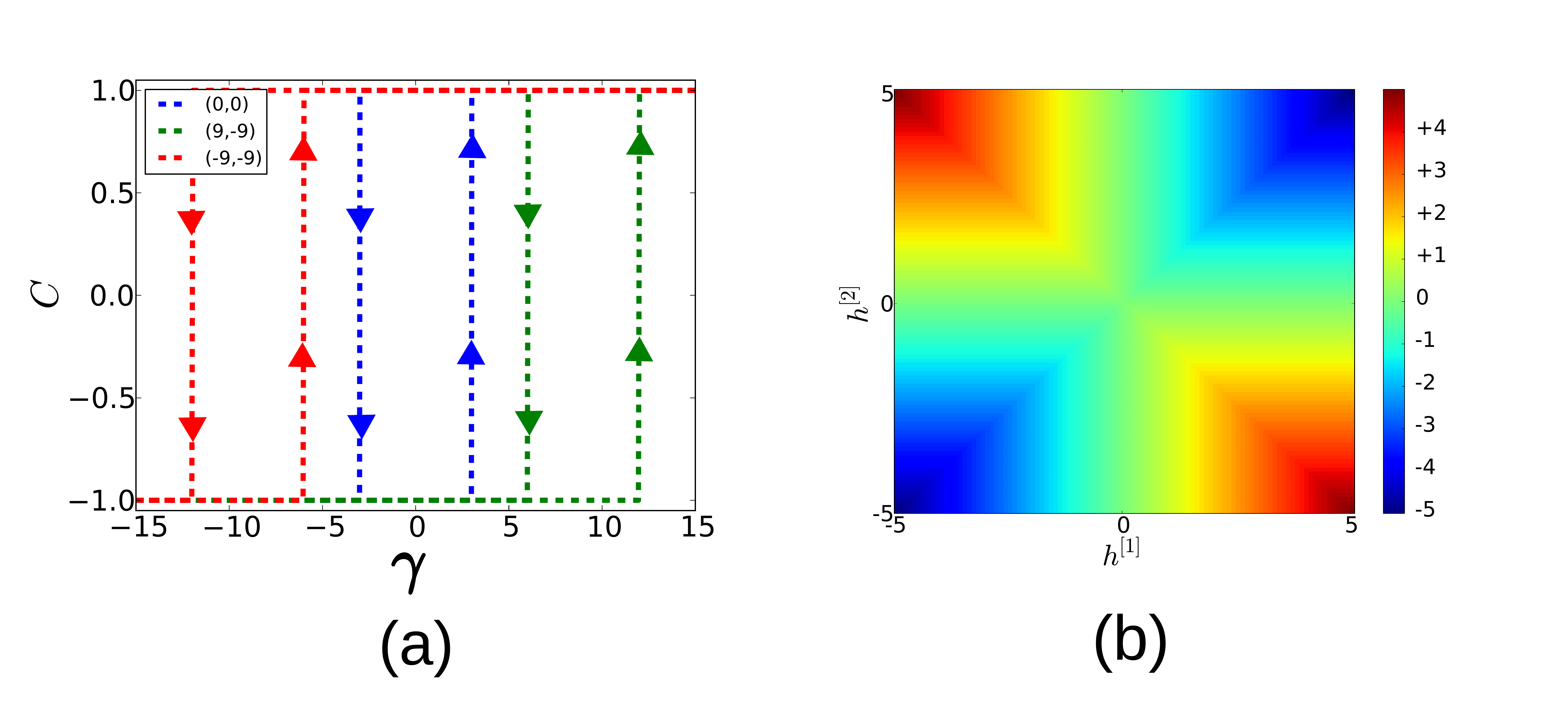}
\caption[]{(a) Plot of the hysteresis cycle of the average
    internal coherence $C$ for different values of ther external
    fields. Even if the qualitative behaviour of the system does not
    depend on the actual values of $h\lay 1 $ and $h\lay 2)$ (i.e.,
    the system is characterised by a sharp transition in $C$ with a
    marked hysteresis loop, whose width is determined solely by the
    average degree on the two layers $\avg{k}$), the exact positions
    of the transition points $\gamma_{+}$ and $\gamma_{-}$ change
    according to the intensity and sign of the external fields. (b)
    Simulated values of $(\gamma_{+}+\gamma_{-})/2$ as a function of
    $h\lay1$ and $h\lay 2$. These numerical results support the
    validity of the empirical relation of Eq.~\eqref{eq:gamma}.}
\label{fig:2b}
\end{figure*}

As suggested before, these qualitative patterns are robust with respect to the strengths of the external fields, which only
determine the exact transition points $\gamma_{+}$ and
$\gamma_{-}$, as shown in Fig.~\ref{fig:2b}(a). We find that the transitions points $\gamma_{+}$ and $\gamma_{-}$ where the hysteresis
loop starts and ends respectively are given by the following empirical non-linear relation:
\begin{equation}
\gamma_{\pm}=\pm \left\langle k \right\rangle/2 - \sign(h\lay 1 h\lay
2)\min\left(|h\lay 1|, |h\lay 2|\right),
\label{eq:gamma}
\end{equation}
where $\left\langle k \right\rangle =
\frac{1}{2N}\sum_{\alpha=1}^{2}\sum_{i,j=1}^{N}a\lay{\alpha}_{ij}$ is the average degree of the two layers. The actual values of $h\lay 1, h\lay 2$ only determine a shift of the metastable region, whereas they do not modify the width of the hysteresis cycle. We support this conjecture by showing in Fig.~\ref{fig:2b}(b) the values of $(\gamma_{+}+\gamma_{-})/2$ (i.e. the center of the hysteresis cycle) obtained from the simulations as a function of $h \lay 1$ and $h \lay 2$, confirming the validity of the relation expressed in Eq.~\eqref{eq:gamma}.

\begin{figure*}[!tb]
\centering
\includegraphics[width=6.1in]{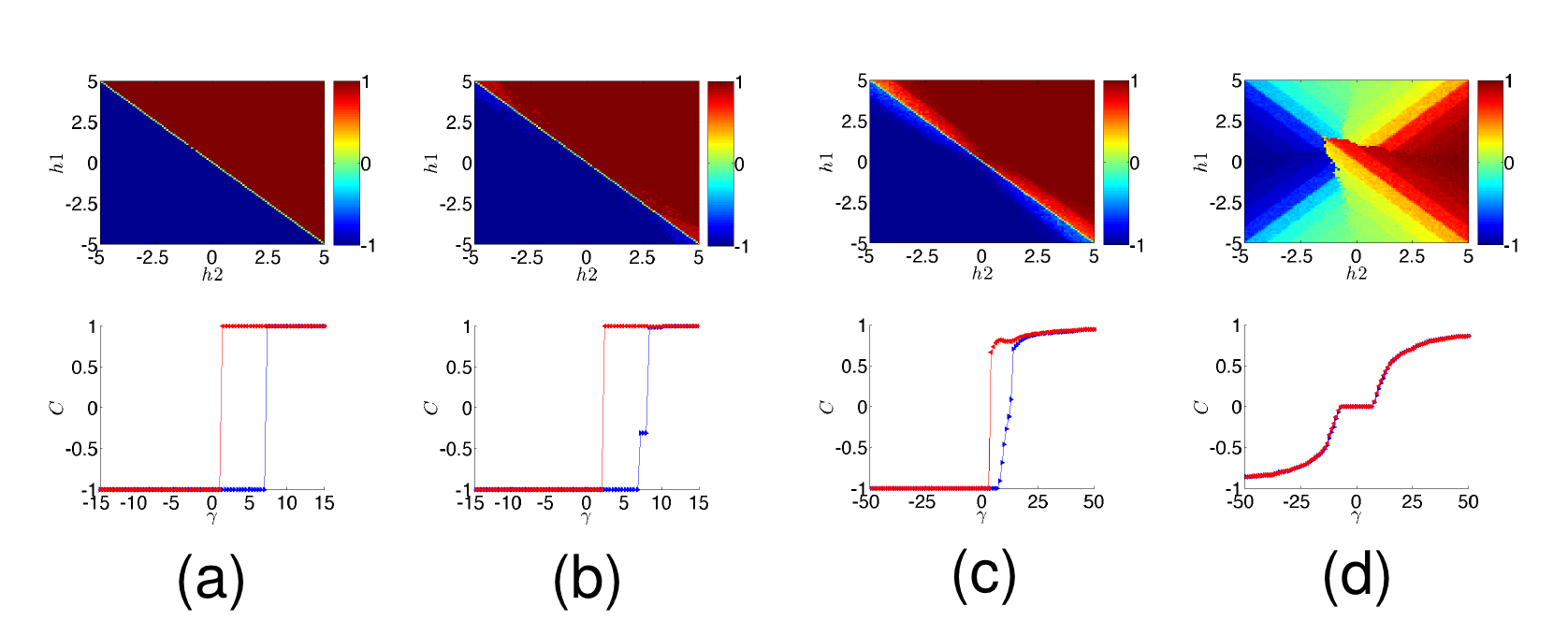}
\caption[]{Phase diagrams of $M\lay 2$ (consensus on the second layer) with respect to the external fields (top panels) for a population of coherent agents, i.e., for $\gamma$ large enough to have average coherence $C=1$ ($\gamma=15$ for panels (a-b) and $\gamma=50$ for panels (c-d)), and the corresponding transition of $C$ (bottom
 panels) as a function of $\gamma$ (forward/backward branch of the hysteresis cycle respectively for blue/red lines) for $h\lay 1=5$, $h\lay 2=-3$. The inter-layer coupling $\chi_i$ for each panel is sampled
 from different distributions, namely (a) $\chi_1=1 \,\, \forall i$, (b)
 half of the agents with $\chi_i=1$ and the remaining half with
 $\chi_i=1/2$, (c) $\chi_i$ uniformly distributed in $[0,1]$, and (d)
 $\chi_i$ uniformly distributed in $[-1,1]$. 
We recall that in this regime, where $C=1$, we have $M\lay 1=M\lay2$.}
\label{fig:3}
\end{figure*}

We conclude that in the case of homogeneous agents the system always
reaches configurations of full consensus on both layers, where the
dominant opinion on each layer is determined by the sign of the
strongest external field (phase diagram in Fig.~\ref{fig:3}, top panel
a). The only exception is given by the critical line $h_1=-h_2$ where
we find $M \lay 2 \approx 0$. As expected, the assumption of
homogeneity of the agents imposes a strong constraint on the dynamics
of the model, leading only to unrealistic patterns of perfect (or null
in the specified particular case) consensus always accompanied by
perfect coherence, but not allowing intermediate configurations. These
sharp scenarios are different from those observed for real-world
systems, where states of partial consensus are often observed. The
heterogeneity in the relative weight and sign assigned to
internal coherence indeed plays a crucial role by favoring the
influence of the mass media over the attainment of full coherence in
the population. Thus, we expect that the relaxation of the homogeneity
hypothesis in our model could lead to milder patterns, with different
levels of consensus at equilibrium, thus better resembling the
observed dynamics of real-world societies.

\subsection{\label{SubSec:B}Heterogeneous agents and the emergence of partial consensus states}
We here consider the case of a population of heterogeneous agents, i.e. $\chi_i$ may be different for each agent $i$. As in the previous section, we set $T=0$, meaning that we neglect the effect of social noise. Such realistic scenarios break the steep transition of the average internal coherence $C$ and allow for the emergence of states of partial consensus in populations of coherent agents. We support this claim by reporting in Fig.~\ref{fig:3} both the phase diagrams of the consensus on the second layer $M \lay 2$ for $\gamma \gg 1$, or equivalently $C=1$, (top panels) and the plot of $C$ as a function of $\gamma$ for a typical choice of the external fields ($h\lay 1=5$, $h\lay 2=-3$ specifically) for a few simple but explanatory cases. 

We first consider the simplest possible setup where half of the
population is assigned $\chi_i=1/2$, whereas the other one is assigned
$\chi_i=1$ [Fig.~\ref{fig:3}(b)], meaning respectively that $50\%$ of
the population is flexible with respect to internal consensus
($\chi_i=1/2$) while the remaining agents are intransigent
($\chi_i=1$). Even if in this case the phase diagram looks similar to
the one in Fig.~\ref{fig:3}(a) for a population of homogeneous agents,
we can already observe the emergence of states of partial consensus
close to the diagonal, i.e., for $|h\lay1|,|h\lay 2|>2.5$. The
breaking of the steep transition in $C$ is also confirmed in the
bottom panel of Fig.~\ref{fig:3}(b).

We then consider in Fig.~\ref{fig:3}(c) the case of an heterogeneous population with $\chi_i\in U(0,1)$, i.e., uniformly distributed in the interval $[0,1]$. In this case, the qualitative behaviour of both $M \lay 2$ and $C$ is similar to the one shown in Fig.~\ref{fig:3}(b). However, as expected due to the increase of the level of heterogeneity of the population, the regions of partial consensus are wider and characterized by lower values of $M \lay 2$ with respect to the previous case. 

Thus, we may expect to find even richer phase diagrams and smoother
transitions in $C$ with respect to the cases presented before if we
further increase the heterogeneity of the population. Indeed, when
$\chi_i$ is sampled uniformly in $[-1, 1]$, the phase diagram looks
qualitatively different: $M \lay 2$ smoothly increases from $-1$ to
$+1$ for increasing values of $h \lay 2$ and fixed $h \lay
1$. Furthermore, the consensus attained in the region $|h\lay 2|<2.5$
with $|h\lay 1|>2.5$ is significantly smaller than in the other
cases. These results suggest that one can smoothly tune the level of
consensus on each topic by choosing the relative strength of the media
acting on the two layers, and yet obtain states in which the majority
of the agents are internally coherent. We also recall that in all the
non-homogeneous cases (Fig.~\ref{fig:3}(b-d)) the system reaches full
coherence, but the transition is not sharp. We conclude by
highlighting that our model, even if simplified, is nevertheless able
to generate non trivial states of partial consensus across the layers
due to the driving effect of mass media, while at the same time
ensuring that each agent will still find itself coherent.  

\begin{figure*}[!tb]
\begin{center}
  \includegraphics[width=6.2in]{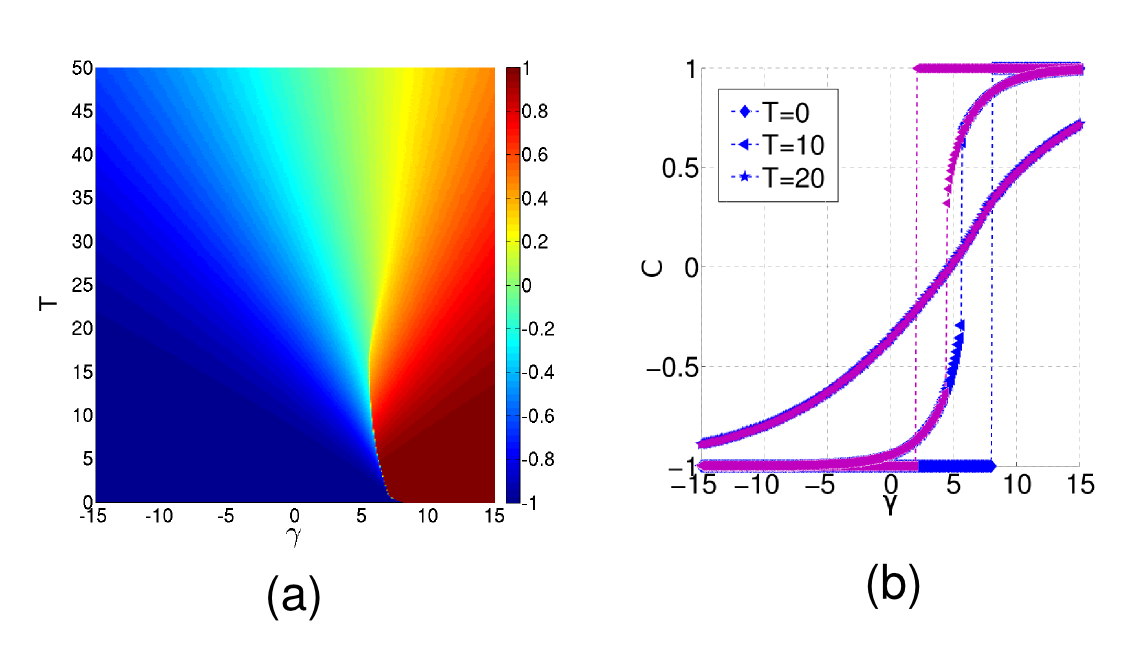}
\caption[]{(a) Average internal coherence $C$ as a function of the noise $T$ and of the global parameter $\gamma$ for $h \lay 1=10$, $h \lay 2=-5$. Simulations are obtained by increasing $\gamma$ adiabatically for fixed $T$ (forward branches shown). There exists a critical value $T_c$ of $T$ below which the system always attains
full coherence ($C=1$) and consensus ($M\lay 1=M\lay 2=+1$) through sharp transitions (consensus not shown). For $T>T_c$ instead the noise becomes dominant and the agents remain incoherent for any $\gamma$ finite. Eventually for $\gamma\rightarrow\infty$ full coherence is obtained via a smooth transition. (b) Projections of $C$ for both increasing and decreasing values of $\gamma$ (dark/light color respectively) for three different values of $T$ (symbols). The transition
 between coherent and incoherent states smoothens as $T$ increases, eventually becoming continuous when $T>T_c \approx 10$.}
 
\label{fig:4}
\end{center}
\end{figure*}

\subsection{\label{SubSec:C}The effect of social noise}
We here consider the case with social noise, i.e. $T >0$. For simplicity, we investigate its effect in a population of homogeneous agents ($\chi_i=1$ $\forall i$). We find that the system exhibits the same qualitative behaviour described in the case $T=0$ for all temperatures below a non-null critical temperature $T_c$, whereas for $T \geq T_c$ it does have absorbing states for finite values of $\gamma$, thus lying in a paramagnetic phase dominated by noise.  

This is shown in Fig.~\ref{fig:4}(a) where we plot $C$ as a function
of both $T$ and $\gamma$ (forward branch of the hysteresis
  cycle) for an exemplary choice of the external fields $h\lay 1$,
$h\lay 2$ with opposite signs. We note that different values of $|h
\lay 1|$, $|h \lay 2|$ do not change qualitatively the results
presented. Indeed, for $T<T_c$ the system exhibits steep transitions
to states of full coherence and consensus. However, when $T$
increases, i.e. as the noise becomes stronger, the jump of the
transition becomes less pronounced and the hysteresis cycle shrinks
considerably, eventually disappearing at $T= T_c$. For $T>T_c$ only
states of partial coherence and consensus can be obtained, and $|C|
\simeq 0$ for $T \gg T_c $. Only in the limit $\gamma \to \infty$, the
population is able to recover full coherence.

This scenario is confirmed by Fig.~\ref{fig:4}(b), where we report
projections of the phase diagram of Fig.~\ref{fig:4}(a) for different
exemplary values of $T$. For $T=0$ the hysteresis cycle is wide and
the jump in $C$ goes from $-1$ to $1$. For $T=10$, slightly below
$T_c$, the hysteresis cycle has almost disappeared and the jump in $C$
is consistently reduced, though still present. For $T=20$, i.e. beyond
the critical level of noise, the transition in $C$ becomes
  continuous and the hysteresis loop disappears. We note that the
noise similarly affects the system in the case of a population of
heterogeneous agents, such that a paramagnetic phase appears beyond
$T_c$ also in this case. Furthermore, we stress that $T_c$ depends non
trivially on the set of parameters of the system. However, deriving
such functional relation is beyond the scope of the present work.

We conclude by recalling that opinion evolution in real social systems
is inevitably affected by noise as already suggested by recent works
on the subject (see for instance ~\cite{drift}). In this section, we
have shown that the behavior of the system for $T=0$ does not change
qualitatively in the presence of noise below some critical value $T_c$
for both a population of homogeneous agents and one of heterogeneous
agents. This ultimately suggests that our finding that heterogeneity
is necessary in population of coherent agents in order to exhibit
realistic states of partial consensus, found for noise-free setups of
our model, may still be relevant for real social systems.

\section{Discussion}
Understanding the elementary mechanisms responsible for the emergence
of consensus in social systems is a fascinating problem that has
stimulated research in several different fields, from sociology to
mathematics, from computer science to theoretical physics, for more
than a few decades. Nevertheless, traditional models used in the field to describe such systems are still far from capturing the essence of the dynamics of real societies.

Indeed, these models of opinion formation overall underestimate the importance of both (i) the existence of many different contexts where social dynamics may develop, and (ii) the variety of interaction patterns that naturally forms between individuals at each of these different aspects. In details, these models are usually based on the simplifying assumption that the social interactions underpinning consensus are essentially homogeneous, whereas real-world societies are instead intrinsically multilayered and multifaceted, meaning that individuals normally interact with several different neighbourhoods in a number of different yet correlated contexts.
Such multilayered structure of social interactions also naturally imply that relationships among each individuals' opinions on many different topics or subjects may exists, thus playing a major role in the formation of an agent's public profile. However, this issue has rarely been addressed in the literature to our knowledge. Overall, these properties of real social systems, force agents to
pursue a balanced trade-off between their internal tendency towards
providing a coherent image of themselves, corresponding to a coherent
set of opinions over the range of contexts in which their social
activities develop, and the external pressure towards local
homogenization that comes from their concurrent participation to
different social circles.

In this work we address the issues (i-ii) thoroughly, and
  propose a novel, yet simple, model of social opinion dynamics which
  is capable to account for them all. Our model is obtained by
  suitably readapting the framework of multilayer networks, which has
  been developed in the last years in different contexts. Remarkably,
  the proposed model suggests that the delicate equilibrium between
  internal agent coherence and responsiveness to external social
pressure in a multilayered social environment might indeed be
one of the fundamental ingredients responsible for the appearance of
non-trivial consensus patterns, such as states of partial
  consensus emerging from a population of coherent agents. Despite
being straightforward in its formulation and relying on rather simple
assumptions, the model we proposed allows to take appropriately into
account the interplay between each agent's tendency towards coherence,
the neighborhood's tendency towards local consensus and the pulling
external forces represented by the persistent action of mass
media. One of the most interesting findings of the present work is
that the introduction of mild heterogeneity in the agents' response to
social pressure fosters the emergence of non-trivial states in which
internal agent's coherence is always reached at the expenses of a
lower level of global consensus. This picture is consistent with what
is widely observed in structured societies~\cite{Baldassarri2008},
where a perfect global consensus is never stable while individuals
tend to adhere to pre-defined sets of social values which they
consider coherent.

Another remarkable effect reproduced by our model is the impact of
mass media pressure, especially in the case where the population is
heterogeneous. In particular, it is interesting to observe that by an
appropriate tuning of the relative strength of the two external fields
representing mass media one can indeed set any desired value of
consensus on each layer, with the possibility of driving the
population from incoherent to more coherent configurations in a
continuous way.  Finally, the results of the study of the role played
by the presence of noise are compatible with real-world scenarios, in
which incomplete or inaccurate information about the state of peers is
the norm and not an exception.

We highlight that the model discussed in this work is limited
  to a specific setting, where both the social and mass-media pressure
  are considered only as a mean-field effect. These assumptions imply
  that the response of agents to both external fields and interactions
  with his/her neighbors is homogeneous, which is only a first-order
  approximation of the real effects of mass media and social pressure
  on a population of agents. A more realistic approach would require
  to consider each agent's adaptive response to such influence, i.e.,
  by both considering that the effect of external field on layer
  $\alpha$ on each node $i$ is a random variable $h_i \lay \alpha$
  drawn from a certain distribution, and considering an
  agent-dependent response to interactions with other individuals,
  i.e. by replacing $J$ with an agent-dependent parameter
  $J_i$. However, we purposedly decided to leave the investigation of
  these generalizations to a future work.

In conclusion, we find it quite intriguing that by taking into account
the presence of concurrent interactions on a variety of different
topics we were able to provide a simple explanation for the formation
of growing patterns of consensus, whose level
appears to be dependent on the strength of mass media pressure, as
long as the agents acknowledge different couplings between their
opinions on the different topics. We believe that the results
presented in this work will spur further research towards a better
understanding of the implications of interconnected and multilayered
interaction patterns on the spreading of opinions and emergence of
consensus in real-world social systems.

\begin{acknowledgments}          
F.B., V.N. and V.L. acknowledge support from the Project LASAGNE,
Contract No.318132 (STREP), funded by the European Commission. This
research utilized Queen Mary's MidPlus computational facilities,
supported by QMUL Research-IT and funded by EPSRC grant EP/K000128/1.
\end{acknowledgments}          


\end{document}